\documentclass[final,5p,times,twocolumn]{elsarticle}

\usepackage{amsthm, amsmath, graphicx}

\usepackage{array}
\usepackage{multirow}
\usepackage{algorithm}
\usepackage{algpseudocode}
\usepackage{color}
\usepackage{hyperref}

\algblock[MyInitBlock]{StartInit}{EndInit}
\algrenewtext{StartInit}{\textbf{Initialize:}}%
\algnotext[MyInitBlock]{EndInit}

\algblock[MyDefineBlock]{StartDefine}{EndDefine}
\algrenewtext{StartDefine}{\textbf{Define:}}%
\algnotext[MyDefineBlock]{EndDefine}

\algnewcommand\algorithmicto{to}
\algrenewtext{For}[3]%
{\algorithmicfor\ $ #1 \gets #2$ \algorithmicto\ $#3$ \algorithmicdo}

\floatstyle{plain}

\newcolumntype{x}[1]{%
>{\centering\hspace{0pt}}p{#1}}%

\DeclareMathOperator*{\argmin}{arg\,min}
\DeclareMathOperator*{\argmax}{arg\,max}

\usepackage{lipsum}
\makeatletter
\def\ps@pprintTitle{%
 \let\@oddhead\@empty
 \let\@evenhead\@empty
 \def\@oddfoot{Accepted for publication in Digital Signal Processing. \href{http://dx.doi.org/10.1016/j.dsp.2013.05.007}{http://dx.doi.org/10.1016/j.dsp.2013.05.007} \hfill\today}%
 \let\@evenfoot\@oddfoot}
\makeatother

\begin{document}

\begin{frontmatter}
\title{Compressed Sensing Signal Recovery via Forward-Backward Pursuit}

\cortext[cor]{Corresponding author}
\author[bte,sabanci]{Nazim Burak Karahanoglu\corref{cor}}
\ead{burak.karahanoglu@tubitak.gov.tr}

\author[sabanci]{Hakan Erdogan}
\ead{haerdogan@sabanciuniv.edu}

\address[bte]{Advanced Technologies Research Institute, TUBITAK-BILGEM, Kocaeli 41470, Turkey}
\address[sabanci]{Department of Electronics Engineering, Sabanci University, Istanbul 34956, Turkey}

\begin{abstract}
Recovery of sparse signals from compressed measurements constitutes an $\ell_0$ norm minimization problem, which is unpractical to solve. A number of sparse recovery approaches have appeared in the literature, including $\ell_1$ minimization techniques, greedy pursuit algorithms, Bayesian methods and nonconvex optimization techniques among others. This manuscript introduces a novel two stage greedy approach, called the Forward-Backward Pursuit (FBP). FBP is an iterative approach where each iteration consists of consecutive forward and backward stages. The forward step first expands the support estimate by the forward step size, while the following backward step shrinks it by the backward step size. The forward step size is larger than the backward step size, hence the initially empty support estimate is expanded at the end of each iteration. Forward and backward steps are iterated until the residual power of the observation vector falls below a threshold. This structure of FBP does not necessitate the sparsity level to be known \textit{a priori} in contrast to the Subspace Pursuit or Compressive Sampling Matching Pursuit algorithms. FBP recovery performance is demonstrated via simulations including recovery of random sparse signals with different nonzero coefficient distributions in noisy and noise-free scenarios in addition to the recovery of a sparse image.

\end{abstract}

\end{frontmatter}

\section{Introduction}
Despite the conventional acquisition process which captures a signal as a whole prior to dimensionality reduction via transform coding, Compressed Sensing (CS) aims at acquisition of sparse or compressible signals directly in reduced dimensions. Mathematically, the ``compressed'' observations are obtained via an observation matrix $\mathbf{\Phi}$
\begin{equation}
\label{Eq1}
\mathbf{y}=\mathbf{\Phi}\mathbf{x},
\end{equation}
where $\mathbf{x}$ is a $K$-sparse signal of length $N$, $K$ is the number of nonzero elements in $\mathbf{x}$, $\mathbf{y}$ is the observation vector of length $M$, and $\mathbf{\Phi}$ is a $M{\times}{N}$ random matrix with $K<M<N$. Once $\mathbf{y}$ is observed, the goal of CS is to recover $\mathbf{x}$, which is analytically ill-posed following the dimensionality reduction via $\mathbf{\Phi}$. Exploiting the sparse nature of $\mathbf{x}$, CS reformulates (\ref{Eq1}) as a sparsity-promoting optimization problem
\begin{equation}\label{Eq:L0Minimization}
\mathbf{x}=\argmin\|\mathbf{x}\|_{0} \;\;\; \text{subject to} \;\;\; \mathbf{y}=\mathbf{\Phi}\mathbf{x},
\end{equation}
where $\|\mathbf{x}\|_0$, called the $\ell_0$ norm by abuse of terminology, denotes the number of nonzero elements in $\mathbf{x}$. As direct solution of (\ref{Eq:L0Minimization}) is computationally intractable, a number of alternative and approximate solutions have emerged in the literature. An overview of mainstream methods is available in \cite{Tropp:CompMeth}, which broadly categorizes CS algorithms as convex relaxation techniques, greedy pursuits, Bayesian methods and nonconvex optimization techniques. Theoretical exact recovery guarantees have also been developed mainly under the Restricted Isometry Property (RIP) \cite{Candes:DecLP, Candes:NOptRec, Candes:RIP} for some of the algorithms. RIP also provides a basis for understanding what type of observation matrices should be employed. Random matrices with Gaussian or Bernoulli entries, or matrices randomly selected from the discrete Fourier transform satisfy RIP with high probabilities \cite{Candes:DecLP, Candes:NOptRec}.

Convex relaxation methods \cite{Candes:DecLP, Candes:PracRec, Donoho:CS, Donoho:OptSparse, Chen:BP, Kim:IPM} replace the $\ell_{0}$ minimization in (\ref{Eq:L0Minimization}) with its closest convex approximation, the $\ell_1$ minimization. Following this modification, recovery is tractable via convex optimization algorithms such as linear programming, as proposed by Basis Pursuit (BP) \cite{Chen:BP}, which is historically the first convex relaxation algorithm. Greedy pursuit algorithms, such as Matching Pursuit (MP) \cite{Mallat:MP}, Orthogonal MP (OMP) \cite{Pati:OMP}, Compressive Sampling MP (CoSaMP) \cite{Needell:CoSAMP}, Subspace Pursuit (SP) \cite{Dai:SP} and Iterative Hard Thresholding (IHT) \cite{Blumensath:IHT1,Blumensath:IHT2}, employ iterative greedy mechanisms.
In addition, \cite{Maleki:TST} provides a framework called Two Stage Thresholding (TST), into which algorithms such as SP and CoSaMP fall.

The manuscript at hand proposes a two stage iterative greedy algorithm, called the Forward-Backward Pursuit (FBP). As the name indicates, FBP employs forward selection and backward removal steps which iteratively expand and shrink the support estimate of $\mathbf{x}$.
With this structure, FBP falls into the general category of TST-type algorithms, while iterative expansion of the support estimate is investigated for the first time in this concept.
Despite their similar structures, FBP has a major advantage over SP and CoSaMP: Since its forward step is larger than the backward one, FBP iteratively expands the support estimate, removing the need for an \textit{a priori} estimate of $K$, which is mostly unknown. Additionally, the backward step of FBP can remove some possibly misplaced indices from the support estimate, which is an advantage over forward greedy algorithms such as OMP. In parallel,
the simulation results in this manuscript demonstrate that FBP can perform better than SP, OMP and BP in most scenarios, which indicates that SP is not necessarily the globally optimum TST scheme as proposed in \cite{Maleki:TST}.

A forward-backward greedy approach for the sparse learning problem, FoBa, has been investigated in \cite{Zhang:FoBa_TransIT}. Though both FoBa and FBP consist of iterative forward and backward steps, the algorithms have some fundamental differences:
FoBa employs strict forward and backward step sizes of one.
On the contrary, the forward step size of FBP is greater than one, while the backward step size might also be.
By increasing the difference between the forward and backward step sizes, FBP terminates in less iterations.
Second, FoBa takes the backward step after a few forward steps based on an adaptive decision. FBP employs no adaptive criterion for taking the backward step, which immediately follows each forward step (Note that using an adaptive criterion is not trivial when the step sizes are greater than one). Finally, FoBa has been applied for the sparse learning problem, while we propose and evaluate FBP for sparse signal recovery from compressed measurements.

This manuscript is organized as follows: First, we give a brief overview of greedy pursuit algorithms. The FBP algorithm is introduced in Section~\ref{sec:fbp}. Section~\ref{sec:results}  demonstrates FBP recovery performance in comparison to the BP, SP and OMP algorithms via simulations involving sparse signals with different nonzero coefficient distributions, phase transitions, noiseless and noisy observations, and a sparse image. We conclude with a brief summary in Section~\ref{sec:conc}.
A preliminary version of this work, \cite{FBP:EUSIPCO}, has been presented at EUSIPCO'2012.

\section{Greedy Pursuits}
\label{sec:greedy}
In this section, we summarize OMP, SP and TST, which are important for our purposes because of their resemblance to the FBP algorithm.
Beforehand, we define the notation that is used throughout the paper: $\mathcal{T}^k$ denotes the estimated support of $\mathbf{x}$ after the $k$th iteration, while $\tilde{\mathcal{T}}^{k}$ stands for the expanded support after the forward selection step of the $k$th iteration. $\mathbf{\tilde{y}}^{k}$ is the approximation of $\mathbf{y}$ after the $k$th iteration and $\mathbf{r}^{k}$ is the residue of $\mathbf{y}$ after the $k$th iteration.
$\mathbf{\Phi}_{\mathcal{J}}$ denotes the matrix consisting of the columns of $\mathbf{\Phi}$ indexed by $\mathcal{J}$, and $\mathbf{x}_{\mathcal{J}}$ is the vector of the elements of $\mathbf{x}$ indexed by $\mathcal{J}$. Finally, $\mathbf{\Phi}^*$ stands for the conjugate of the matrix $\mathbf{\Phi}$. Note that each column of $\mathbf{\Phi}$ is sometimes referred to as an atom in the rest.

OMP is a forward greedy algorithm that searches for the support of $\mathbf{x}$ by identifying one element per iteration. It starts with an empty support estimate: $\mathcal{T}^{0} = \emptyset$ and $\mathbf{r}^{0} = \mathbf{y}$. At the iteration $k$, OMP expands $\mathcal{T}^{k-1}$ with the index of the dictionary atom closest to $\mathbf{r}^{k-1}$, i.e. it selects the index of the largest magnitude entry of $\mathbf{\Phi}^*\mathbf{r}^{k-1}$. Next, $\mathbf{\tilde{y}}^{k}$ is computed via orthogonal projection of $\mathbf{y}$ onto $\mathbf{\Phi}_{\mathcal{T}^{k}}$ and the residue is updated as $\mathbf{r}^{k} = \mathbf{y} - \mathbf{\tilde{y}}^{k}$. The iterations are carried out until the termination criterion is met.
In this work, we stop OMP when $\|\mathbf{r}^{k}\|_2 \leq \varepsilon\|\mathbf{y}\|_2$ similar to the termination criterion of FBP.

SP and CoSaMP combine selection of multiple columns per iteration with pruning, keeping $K$-element support sets throughout the iterations. At iteration $k$, SP first expands $\mathcal{T}^{k-1}$ with the indices of the $K$ largest magnitude elements of $\mathbf{\Phi}^*\mathbf{r}^{k-1}$, obtaining the extended support $\tilde{\mathcal{T}}^k$ of size $2K$ (Alternatively, CoSaMP expands $\mathcal{T}^{k-1}$ by $2K$ elements.). In the second step, the orthogonal projection coefficients of $\mathbf{y}$ onto $\mathbf{\Phi}_{\tilde{\mathcal{T}}^k}$ are computed, and $\mathcal{T}^k$ is obtained by pruning the indices of all but the $K$ largest magnitude projection coefficients from $\tilde{\mathcal{T}}^k$.
$\mathbf{r}^k$ is finally computed using the approximation $\mathbf{\tilde{y}}^k$ which is obtained by orthogonal projection of $\mathbf{y}$ onto $\mathbf{\Phi}_{\mathcal{T}^k}$.
The iterations are stopped when $\|\mathbf{r}^k\|_2 \geq \|\mathbf{r}^{k-1}\|_2$.
CoSaMP and SP are both provided with RIP-based exact recovery guarantees. On the other hand, they employ equal forward and backward step sizes, which lead to a fixed support size between the iterations. This necessitates an \textit{a priori} estimate of the sparsity level $K$. This is an important handicap in most practical cases, where $K$ is either unknown or it is not desired to fix it.

Recently, Maleki and Donoho have introduced the TST framework \cite{Maleki:TST}, into which algorithms such as SP and CoSaMP fall. TST algorithms employ a two stage iterative scheme which first updates the sparse estimate and then prunes it by thresholding. Next, the optimal coefficients are computed by orthogonal projection of $\mathbf{y}$ onto the pruned support. This is followed by a second thresholding operation that yields a new sparse estimate.
\cite{Maleki:TST} also presents a tuned SP algorithm, which turns out to be the empirically optimal TST scheme for sparse signals with constant amplitude nonzero elements.
This algorithm employs a tuned support size which is decided on-the-fly depending on the pre-computed phase transition curves for the particular $M$ and $N$ values of interest. The motivation behind this tuning is to select the support estimate at least as large as the largest sparsity level SP can exactly recover.
However, as the results in \cite{Maleki:TST} also indicate, choosing the support size of SP larger than the actual sparsity level degrades the recovery performance. Hence, this tuned SP algorithm is subject to perform worse than the oracle-SP, which incorporates the actual sparsity level.

\section{Forward-Backward Pursuit}
\label{sec:fbp}

Forward-backward pursuit is an iterative two stage algorithm.
The first stage of FBP is the forward step which expands the support estimate by $\alpha>1$ atoms, where we call $\alpha$ the forward step size.
These $\alpha$ indices are chosen as the indices of the dictionary atoms which are maximally correlated with the residue, following the motivation of obtaining the best match to it.
Then, FBP computes the orthogonal projection of the observed vector onto the subspace defined by the support estimate.
Next, the backward step prunes the support estimate by removing $\beta<\alpha$ indices with smallest contributions to the projection.
Similar to $\alpha$, we refer to $\beta$ as the backward step size.
The orthogonality of the residue to the subspace defined by the pruned support estimate is ensured by a second projection of the residue onto this subspace.
These forward and backward steps are iterated until the energy of the residue either vanishes or is less than a threshold which is proportional to the energy of the observed vector.

\subsection{The Proposed Method}

The FBP algorithm can now be outlined as follows:  We initialize the support estimate as $\mathcal{T}^{0} = \emptyset$, and the residue as $\mathbf{r}^{0} = \mathbf{y}$. At iteration $k$, first the forward step expands $\mathcal{T}^{k-1}$ by indices of the $\alpha$ largest magnitude elements in $\mathbf{\Phi}^*\mathbf{r}^{k-1}$. This builds up the expanded support set $\tilde{\mathcal{T}}^{k}$. Then the projection coefficients are computed by the orthogonal projection of $\mathbf{y}$ onto $\mathbf{\Phi}_{\tilde{\mathcal{T}}^{k}}$. The backward step prunes $\tilde{\mathcal{T}}^{k}$ by removing the $\beta$ indices with the smallest magnitude projection coefficients. This produces the final support estimate $\mathcal{T}^{k}$ of the $k$th iteration. Finally, the projection coefficients $\mathbf{w}$ for the vectors in $\mathbf{\Phi}_{\mathcal{T}^{k}}$ are computed via the orthogonal projection of $\mathbf{y}$ onto $\mathbf{\Phi}_{\mathcal{T}^{k}}$, and the residue is updated as
$\mathbf{r}^{k} = \mathbf{y} - \mathbf{\Phi}_{\mathcal{T}^{k}}\mathbf{w}$.
The iterations are carried on until  $\|\mathbf{r}^{k}\|_2 < \varepsilon\|\mathbf{y}\|_2$. After termination of the algorithm at the $l$th iteration, $\mathcal{T}^{l}$ gives the support estimate for $\mathbf{x}$, while $\mathbf{w}$ contains the corresponding nonzero values. The pseudo-code of FBP is given in Algorithm~\ref{alg:chp6_FBP}.

\begin{algorithm}[!t]{}
\caption{FORWARD-BACKWARD PURSUIT}
\label{alg:chp6_FBP}
\begin{algorithmic}
\vskip2mm
\State \textbf{input:} $\mathbf{\Phi}$, $\mathbf{y}$
\State \textbf{define:} $\alpha$, $\beta$, $K_{\text{max}}$, $\varepsilon$
\State \textbf{initialize:} $\mathcal{T}^{0} = \emptyset$, $\mathbf{r}^{0} = \mathbf{y}$, $k = 0$
\While{true}
\State $k = k + 1$
\State \textit{forward step:}
\State \hspace{3cm} $\mathcal{T}_f = \argmax\limits_{\mathcal{J}: |\mathcal{J}|=\alpha} \left\|\mathbf{\Phi}^*_\mathcal{J} \mathbf{r}^{k-1}\right\|_1$
\State \hspace{3cm} $\tilde{\mathcal{T}}^{k} = \mathcal{T}^{k-1} \cup T_f$
\State \hspace{3cm} $\mathbf{{w}} = \argmin\limits_{\mathbf{w}} \left\| \mathbf{y} - \mathbf{\Phi}_{\tilde{\mathcal{T}}^{k}}\mathbf{w} \right\|_2$
\State \textit{backward step:}
\State \hspace{3cm} $\mathcal{T}_b = \argmin\limits_{\mathcal{J}: |\mathcal{J}|=\beta} \left\|\mathbf{w}_{\mathcal{J}}\right\|_1 $
\State \hspace{3cm} $\mathcal{T}^{k} = \tilde{\mathcal{T}}^{k} - \mathcal{T}_b$
\State \textit{projection:}
\State \hspace{3cm} $\mathbf{w} = \argmin\limits_{\mathbf{w}} \left\| \mathbf{y} - \mathbf{\Phi}_{\mathcal{T}^{k}}\mathbf{w} \right\|_2$
\State \hspace{3cm} $\mathbf{r}^{k} = \mathbf{y} - \mathbf{\Phi}_{\mathcal{T}^{k}}\mathbf{w}$
\State \textit{termination rule:}
\State \hspace{3cm} \textbf{if} $\left\| \mathbf{r}^{k} \right\|_2 \leq \varepsilon\left\|\mathbf{y}\right\|_2$ or $\left|\mathcal{T}^k\right| \geq K_{\text{max}}$ \textbf{then}
\State \hspace{3.2cm} \textbf{break}
\State \hspace{3cm} \textbf{end if}
\EndWhile
\State $\tilde{\mathbf{x}} = 0$
\State $\tilde{\mathbf{x}}_{\mathcal{T}^{k}} = \mathbf{w}$
\State \Return $\mathbf{\tilde{x}}$
\end{algorithmic}
\end{algorithm}

As for the termination parameter $\varepsilon$, we choose a very small value in practice ($10^{-6}$ for the experiments below) when the observations are noise-free. For noisy observations, $\varepsilon$ should be selected depending on the noise level. To avoid the algorithm running for too many iterations in case of a failure, the maximum size of the support estimate is also limited by $K_{\text{max}}$.
Note that, the specific choice of $K_{\text{max}}$ has no significant effect on the recovery accuracy given it is a bit larger than the underlying sparsity level $K$ (to assure  the correct solution may be found before the support size reaches $K_{\text{max}}$).
Since $K$ cannot be known in practice, we may set $K_{\text{max}}$ either large enough or simply as $K_{\text{max}} = M$.
In addition, the phase transitions of FBP may also be used for obtaining a large enough estimate for $K_{\text{max}}$ given $N$ and $M$ in a specific scenario. That is, given the empirical phase transition curve, $K_{\text{max}}$ can be chosen such that it corresponds to a sparsity ratio which lies over the phase transition curve for the particular $M$ and $N$ values.

An important issue for the performance of FBP is the choice of the forward and backward step sizes. The forward step size $\alpha$ should be chosen larger than 1. It is possible to choose $\alpha$ as large as problem-specific constraints allow, while a reasonable approach would obviously be selecting it small in comparison to the observation length $M$ in order to avoid linearly dependent subsets in the expanded support estimate after the forward step. As for the backward step, $\beta$ should be smaller than $\alpha$ by the definition of FBP, since the support estimate should be enlarged at each iteration.
In order to find an empirically optimal rule for choosing $\alpha$ and $\beta$, we present phase transition curves of FBP with various $\alpha$ and $\beta$ choices among the simulation results below. It turns out that choosing $\alpha \in [0.2M, 0.3M]$ and $\beta=\alpha-1$ leads to the optimal recovery performance in practice, whereas the algorithm is also quite robust to other choices of $\alpha$ and $\beta$ as well. In particular, choosing $\beta<\alpha-1$ speeds up the algorithm without a severe decrement in the recovery accuracy as demonstrated below.

\subsection{Relations to Other Greedy Algorithms} \label{sec:chp6_fbp_relations}

Forward greedy algorithms, such as OMP and other MP variants, which enlarge the support estimate iteratively via forward selection steps, have a fundamental drawback by definition: Since they possess no backward removal mechanism, any index that is inserted into the support estimate cannot be removed. That one or more incorrect elements remain in the support until termination may cause the recovery to fail. FBP, on the contrary, employs a backward step, which provides means for removal of atoms from the support estimate. This gives FBP the ability to cover up for the errors made by the forward step.
To illustrate, consider a well-known example: Let $\mathbf{x}$ be the summation of two equal magnitude sinusoids with very close frequencies, $f_1$ and $f_2$, and $\mathbf{\Phi}$ be an overcomplete sinusoidal dictionary, containing atoms with frequencies $f_1$, $f_2$ and $f_3 = (f_1{+}f_2)/2$ among others. The first iteration of OMP selects the component with frequency $f_3$. Then, during the next iterations, the algorithm tries to cover for this error by choosing components other than the two correct ones and fails.
Instead, assume we run FBP with $\alpha=3$ and $\beta=1$. During the forward step of the first iteration, FBP selects all the three components with frequencies $f_1$, $f_2$ and $f_3$. Following orthogonal projection, the backward step will eliminate $f_3$, and the recovery will be successful after the first iteration\footnote{Note that the success of FBP in this case depends on the choice of $\alpha$ and $\beta$, however, this example still illustrates the motivation behind the backward removal step in a very simple way.}.

In contrast to SP and CoSaMP, the FBP algorithm does not require an \textit{a priori} estimate of the sparsity level $K$. Unlike the tuned TST, it does not necessitate a tuning of the support size either. As explained above, FBP enlarges the support estimate by $\alpha-\beta$ indices at each iteration until termination of the algorithm, which depends on the residual power, and not on the sparsity level. Hence, neither the forward and backward steps nor the termination criterion require an estimate of the sparsity level.
In addition, the forward and backward step sizes of FBP may be chosen proportional to $M$ with a simple empirical strategy as demonstrated below, while the recovery performance is quite robust to this choice. These make the FBP algorithm easily applicable in practice in contrast to SP and CoSaMP. This, however, comes at a cost: The theoretical guarantees cannot be provided in a way similar to SP or CoSaMP, which make use of the support size being fixed as $K$ after the backward step. For the time being, we cannot provide a complete theoretical analysis of FBP, and leave this as future work. Note that, however, most of the theoretical analysis steps of SP or CoSaMP also hold for FBP.

\section{Experimental Evaluation}
\label{sec:results}

This section is reserved for the demonstration of the FBP recovery performance in comparison to BP, SP, OMP.
For this purpose, we run recovery simulations involving different nonzero coefficient distributions, noiseless and noisy observations, and a sparse image.
First, we compare the exact recovery rates, average recovery error and run times of FBP with those of OMP, SP and BP for signals with nonzero elements drawn from the Gaussian and uniform distributions.
In order to generalize the results to a wide range of $M$ and $K$ along with different nonzero element distributions, we provide the empirical phase transition curves, which are obtained using the procedure in \cite{Maleki:TST}.
Meanwhile, these phase transition curves also serve for the purpose of investigating optimal $\alpha$ and $\beta$ choices.
We then demonstrate recovery from noisy observations, and finally test our proposal on a sparse image to illustrate the recovery performance for a realistic coefficient distribution.

Results of the 1D simulations are presented as averages over three different data sets, where the nonzero entries of the test samples in each set are selected from different random ensembles. The nonzero entries of the Gaussian sparse signals are drawn from the standard Gaussian distribution. Nonzero elements of the uniform sparse signals are distributed uniformly in $[-1,1]$, while the constant amplitude random sign (CARS) sparse signals have nonzero elements with unit magnitude and random sign. During the experiments, a different observation matrix $\mathbf{\Phi}$ is drawn from the Gaussian distribution with mean $0$ and standard deviation $1/N$ for each test signal. All experiments are performed in the MATLAB environment. For fair comparison of the run times, algorithms share similar structures.
The tests are run on a modest laptop with Pentium Dual-Core CPU at 2.3 GHz and 2 GB memory under Windows 7.

As for the termination parameters, $\varepsilon=10^{-6}$ in the noise-free case, while it depends on the signal-to-noise ratio (SNR) under noisy conditions.
$K_{\text{max}}$, which is not critical for the recovery performance as discussed in Section~\ref{sec:fbp}, is chosen large enough, particularly either $K_{\text{max}}=M$ or $K_{\text{max}} > M/2$.
Note that the same $\varepsilon$ and $K_{\text{max}}$ are also used for OMP.

\subsection{Exact Recovery Rates and Reconstruction Error}

First, we compare the exact recovery rates, recovery error and run times of FBP using various $\alpha$ and $\beta$ values with those of OMP, SP and BP. In these simulations, the signal and observation sizes are fixed as $N=256$ and $M=100$ while $K$ varies in $[10, 45]$.
For each $K$, recovery simulations are repeated over 500 randomly generated Gaussian and uniform sparse signals. The recovery error is expressed in terms of Average Normalized Mean-Squared-Error (ANMSE), which is defined as
\begin{equation}\label{Eq:ANMSE}
    ANMSE = \frac{1}{500} \sum_{i=1}^{500}{\frac{\|\mathbf{x}_i - \hat{\mathbf{x}}_i\|_2^2}{\|\mathbf{x}_i\|_2^2}}
\end{equation}
where $\hat{\mathbf{x}}_i$ is the recovery of the $i$th test vector $\mathbf{x}_i$. In addition, we present the exact recovery rates, which represent the ratio of perfectly recovered test samples to the whole test data. The exact recovery condition is selected as $\|\mathbf{x}-\mathbf{\hat{x}}\|_2 \leq 10^{-2}\|\mathbf{x}\|_2$ following \cite{Maleki:TST}. In these tests, we select $K_{\text{max}}=55$ to allow for exact recovery of sparse signals up to about $M/2=50$ nonzero elements. Note that, for the specific $N$ and $M$ values in this experiment, the phase transition occurs well below $M/2$ (see the phase transitions below), hence choosing $K_{\text{max}}>M/2$ is sufficient.

\begin{figure*}[!t]
\begin{center}
\includegraphics[width=\linewidth]{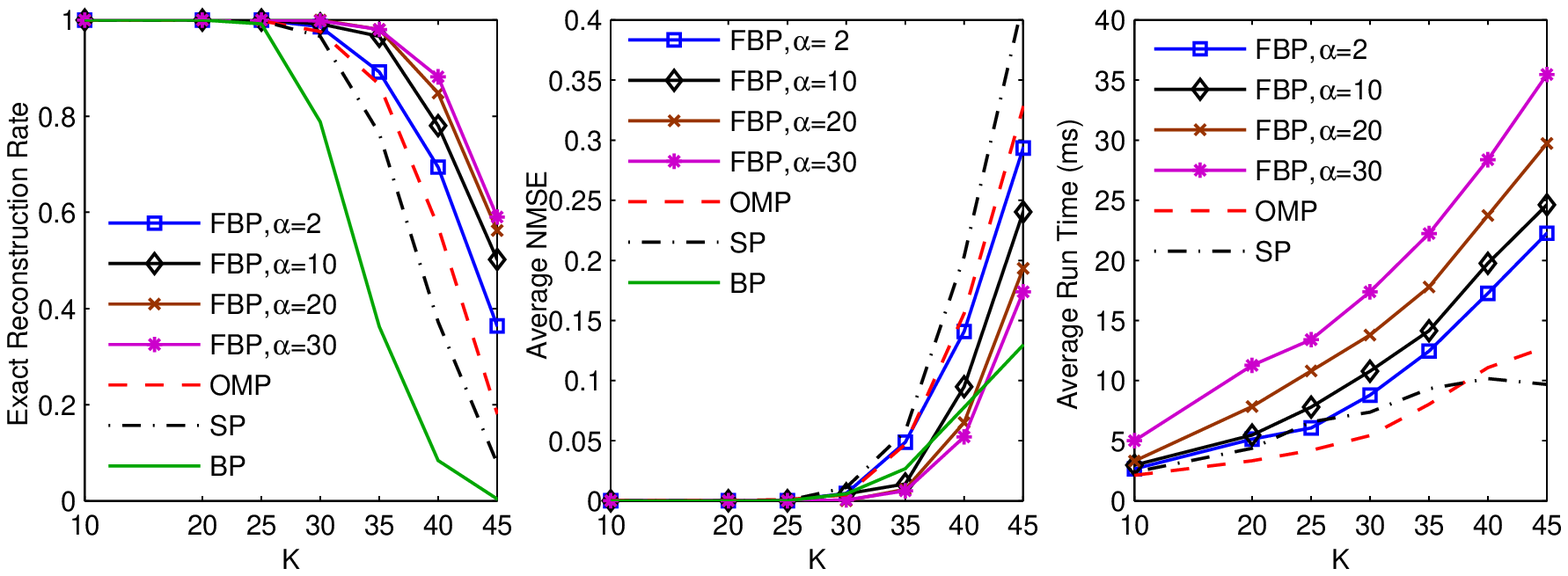}
\end{center}
\caption{Reconstruction results over sparsity for the Gaussian sparse vectors. For FBP, $\beta=\alpha-1$.}
\label{fig:gauss1}
\vspace{8pt}
\begin{center}
\includegraphics[width=\linewidth]{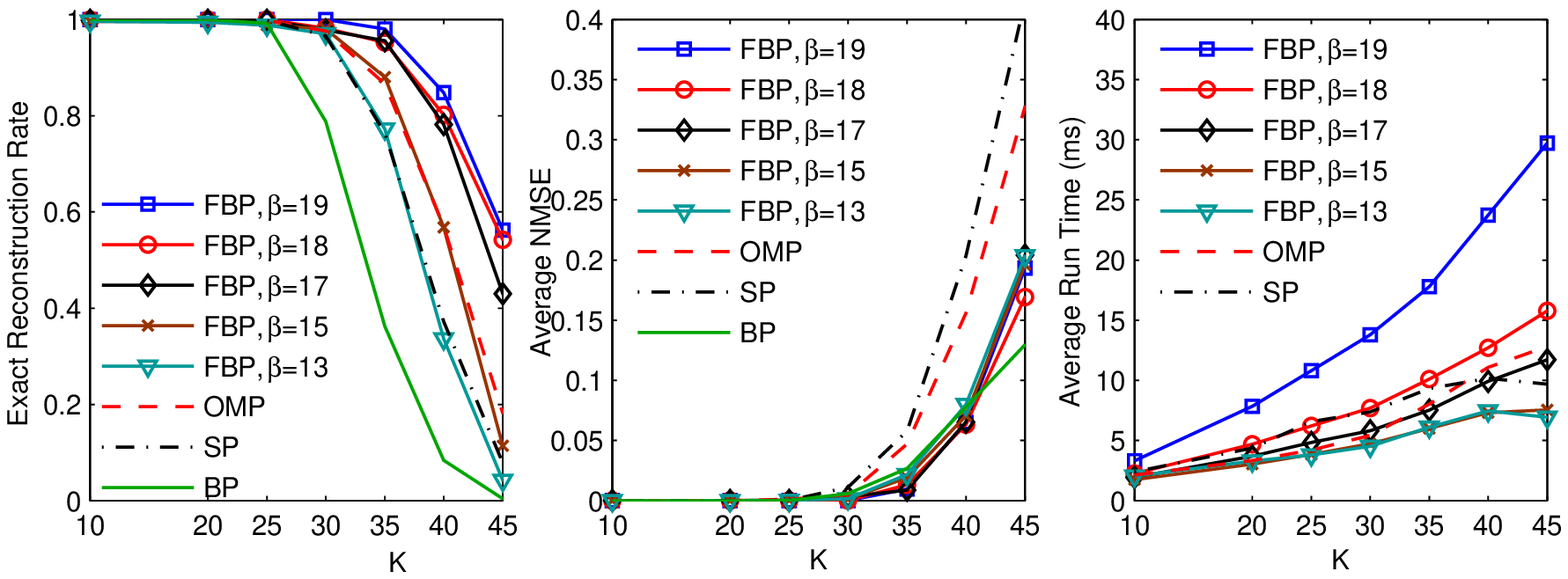}
\end{center}
\caption{Reconstruction results over sparsity for the Gaussian sparse vectors. For FBP, $\alpha=20$.}
\label{fig:gauss2}
\end{figure*}

\begin{figure*}[!t]
\begin{center}
\includegraphics[width=\linewidth]{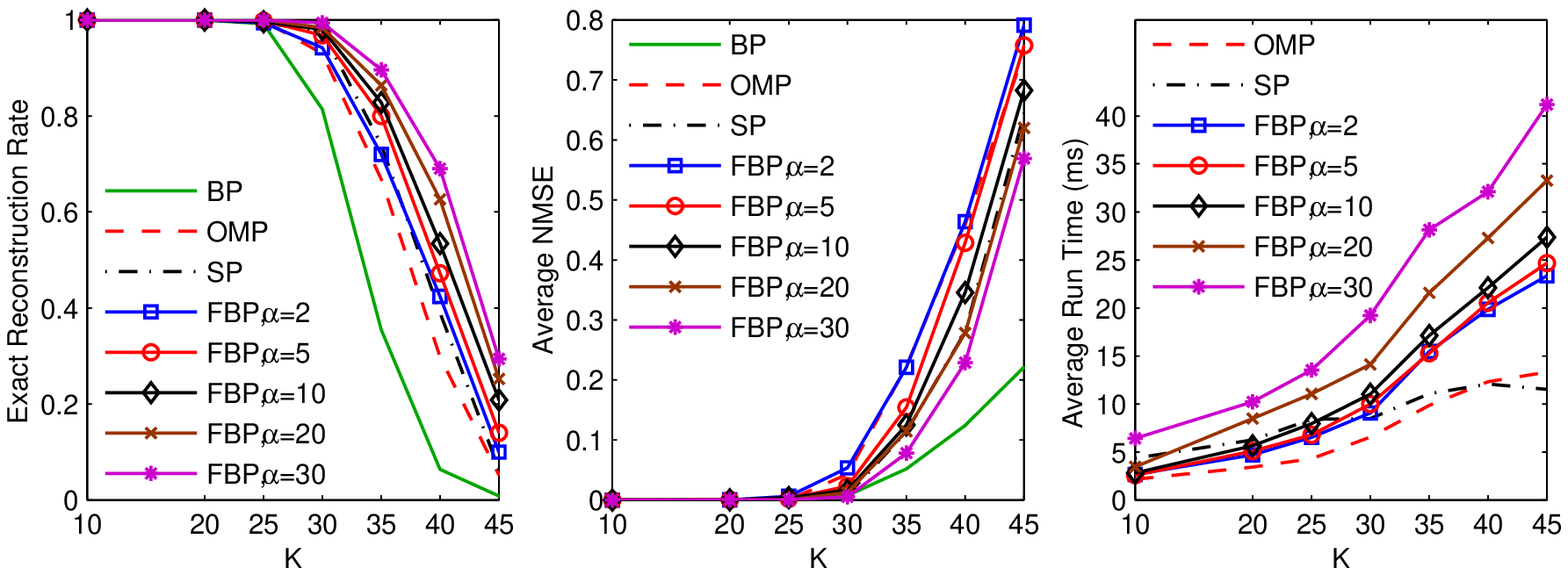}
\end{center}
\caption{Reconstruction results over sparsity for uniform sparse vectors. For FBP, $\beta=\alpha-1$.}
\label{fig:uniform1}
\vspace{8pt}
\begin{center}
\includegraphics[width=\linewidth]{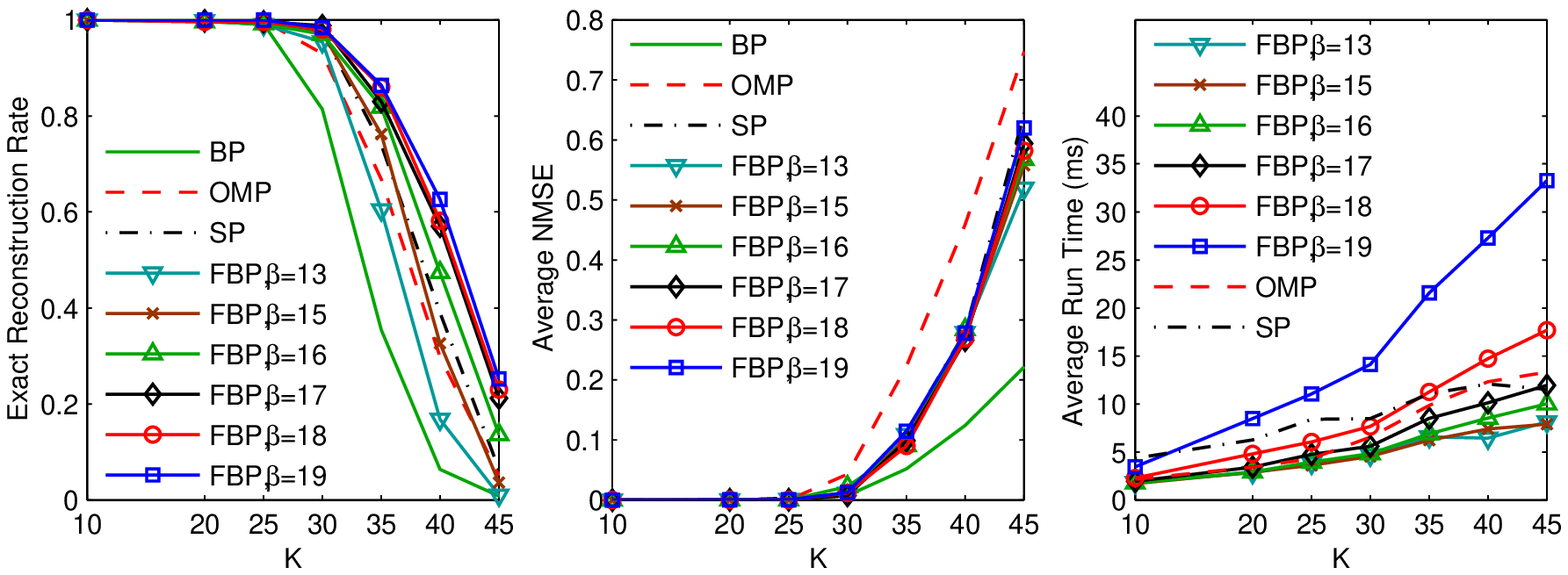}
\end{center}
\caption{Reconstruction results over sparsity for uniform sparse vectors. For FBP, $\alpha=20$.}
\label{fig:uniform2}
\end{figure*}

Fig.~\ref{fig:gauss1} and \ref{fig:gauss2} depict the reconstruction performance of FBP with various $\alpha$ and $\beta$ choices for the Gaussian sparse signals in comparison to OMP, BP and SP.
Fig.~\ref{fig:gauss1} is obtained by varying $\alpha$ in $[2, 30]$, while $\beta=\alpha-1$. That is, the forward step size varies, while the support estimate is expanded by one element per iteration.
For Fig.~\ref{fig:gauss2}, $\alpha$ is selected as 20, and $\beta$ is altered in $[13, 19]$, changing the increment in the support size per iteration for a fixed forward step size.
The run times of the FBP, SP and OMP algorithms are also compared, while BP is excluded as it is incomparably slower than the other algorithms.
Analogous results are provided in Fig.~\ref{fig:uniform1} and \ref{fig:uniform2} for the uniform ensemble as well.

According to Fig.~\ref{fig:gauss1}, increasing $\alpha$ while keeping the support increment $\alpha-\beta$ fixed improves the recovery performance of FBP.
We observe that the exact recovery rates of FBP are significantly better than the other candidates for all choices of $\alpha$, even including the modest choice $\alpha=2$.
BP, SP, and OMP start to fail at around $K=25$, where FBP is still perfect for all choices of $\alpha$. Moreover, for $\alpha \geq 20$, the FBP failures begin only when $K > 30$.
As for ANMSE, FBP is the best performer when $\alpha\geq 20$. With this setting, BP can beat FBP in ANMSE only when $K>40$. In addition, FBP yields better recovery rates than OMP and SP for all choices of $\alpha$.

In Fig.~\ref{fig:gauss2}, we observe that increasing $\beta$ for a fixed $\alpha$ improves the recovery performance.
In this case, the exact recovery rate of FBP significantly increases with the backward step size, while ANMSE remains mostly unaltered.
This indicates that when $\beta$ is increased, nonzero elements with smaller magnitudes, which do not significantly change the recovery error, can be more precisely recovered.
In comparison to the other algorithms involved in the simulation, FBP is the best performer for $\beta > 15$. Similar to the previous test case, BP can produce lower ANMSE than FBP only for $K>40$.

Investigating Fig.~\ref{fig:uniform1} and \ref{fig:uniform2}, which depict recovery results for the uniform sparse signals, we observe a similar behavior as well. FBP yields better exact recovery rates than the other algorithms when $\alpha$ and $\beta$ are large enough, i.e. $\alpha \geq 5$ in Fig.~\ref{fig:uniform1}, and $\beta\geq16$ in Fig.~\ref{fig:uniform2}, while BP can perform better than FBP in terms of the average error when $K>30$.

As for the run times, we expectedly observe that increasing $\alpha$ or $\beta$ slows down FBP. This is due to the decrease in the increment of the support size per iteration, which increases the number of iterations and the number of required orthogonal projection operations. Moreover, the dimensions of the orthogonal projection operations also increase with the forward step size.
On the other hand, increasing $\alpha-\beta$ decreases the number of necessary iterations, as a result of which FBP terminates faster. More important, the run time of FBP, SP and OMP are very close when $\alpha=20$ and $\beta \leq \alpha-2$.
In case $\alpha=20$ and $\beta=17$, the speed of FBP and OMP are almost the same, whereas the exact recovery rate of FBP is significantly better than the other algorithms involved. Note that the speed of FBP can be improved by removing the orthogonal projection after the backward step, at the expense of a slight degradation in the recovery performance.

\subsection{Phase Transitions}

Phase transitions are important for empirical evaluation of CS recovery algorithms over a wide range of the sparsity level and the observation length.
Below, we present the empirical phase transition curves of the FBP algorithm in comparison to those of the OMP, SP, and BP algorithms.
These graphs are obtained from recovery simulations involving 200 Gaussian, uniform or CARS sparse signals each.
Below, we first depict phase transitions of FBP  with different $\alpha$ and $\beta$ choices in order to investigate the optimality of these over the observation length. These simulations provide us an empirical strategy about how to choose the FBP step sizes in relation to $M$. Next, we compare FBP with two different settings to BP, OMP and SP for the three test sets. In this test set, we set $K_{\text{max}}=M$. This choice is reasonable here, since we are interesting in recovering signals with a very wide $K/M$ range.

To explain how we obtain the phase transitions, let us first define normalized measures for the observation length and the sparsity level: $\lambda = M/N$ and $\rho = K/M$.
To obtain the phase transition curves, we keep the signal length fixed at $N=250$, and alter $M$ and $K$ to sample the $\{\lambda, \rho\}$ space for $\lambda \in [0.1, 0.9]$ and $\rho \in (0,1]$.
For each $\{\lambda,\rho\}$ tuple, we randomly generate 200 sparse instances and run FBP, OMP, BP and SP algorithms for recovery.
The exact recovery condition being $\|\mathbf{x}-\mathbf{\hat{x}}\|_2 \leq 10^{-2}\|\mathbf{x}\|_2$ as before, exact recovery rate is obtained for each $\{\lambda,\rho\}$ tuple and each algorithm. The phase transitions are then obtained using the methodology described in \cite{Maleki:TST}. That is, for each $\lambda$, we employ a generalized linear model with logistic link to describe the exact recovery curve over $\rho$, and then find the $\rho$ value which yields $50\%$ exact recovery probability.

Phase transitions provide us important means for finding an empirical way of choosing $\alpha$ and $\beta$ optimally.
As discussed in \cite{Maleki:TST}, the phase transition curve is mostly a function of  $\lambda$. That is, it remains unaltered when $N$ changes. Moreover, the transition region turns out to be narrower with increasing $N$. These claims are also supported by some other publications in the literature \cite{Berinde_OptItRefNo42, Donoho:fastsolution, Donoho:StOMP, Donoho:CountingFaces}.
Hence, in order to find an optimal set of step sizes for FBP, we need to have a look at the phase transitions using different $\alpha$ and $\beta$ parameters.
For a better understanding of their optimality, $\alpha$ and $\beta$ should not be fixed but be proportional to $M$.
Trying to find fixed $\alpha$ and $\beta$ values is subject to fail mainly for very low or very high $\lambda$ values.
In other words, it would not be possible to find a fixed optimal set $\{\alpha,\beta\}$ for the whole $\lambda$ range even when we fix $N$.
This is, however, possible when $\alpha$ is proportional to $M$, and $\beta$ is related to the chosen $\alpha$ value. In order to find an optimal choice, we run two distinct sets of simulations: First, we vary $\alpha$ in $[0.1M,\;0.4M]$, whereas $\beta=\alpha-1$. Then we fix $\alpha=0.2M$, and select $\beta$ either in $[0.7\alpha,\;0.9\alpha]$ or as $\alpha-1$.

\begin{figure*}[!t]
\begin{center}
\includegraphics[width=11.5cm]{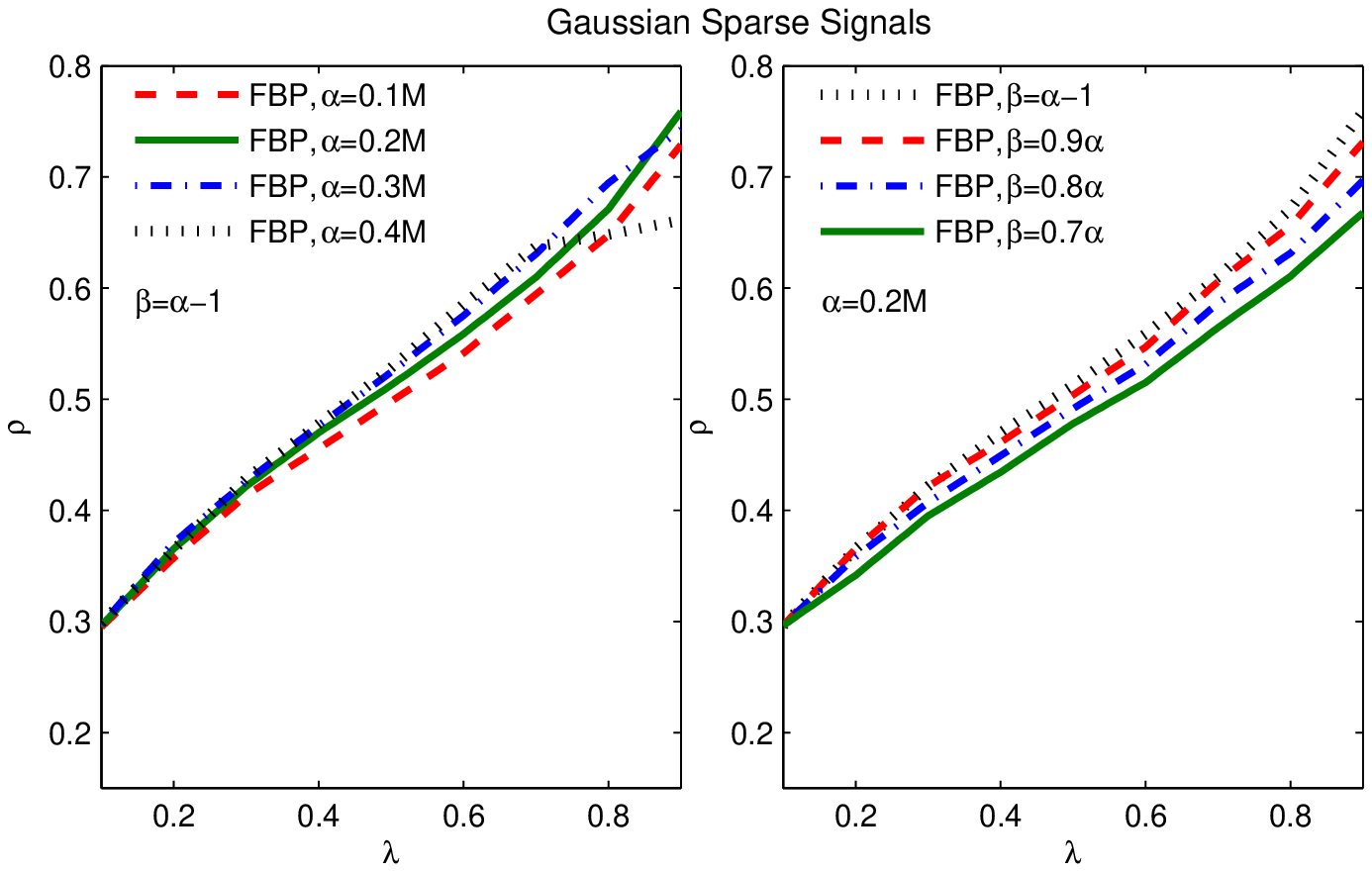}
\includegraphics[width=11.5cm]{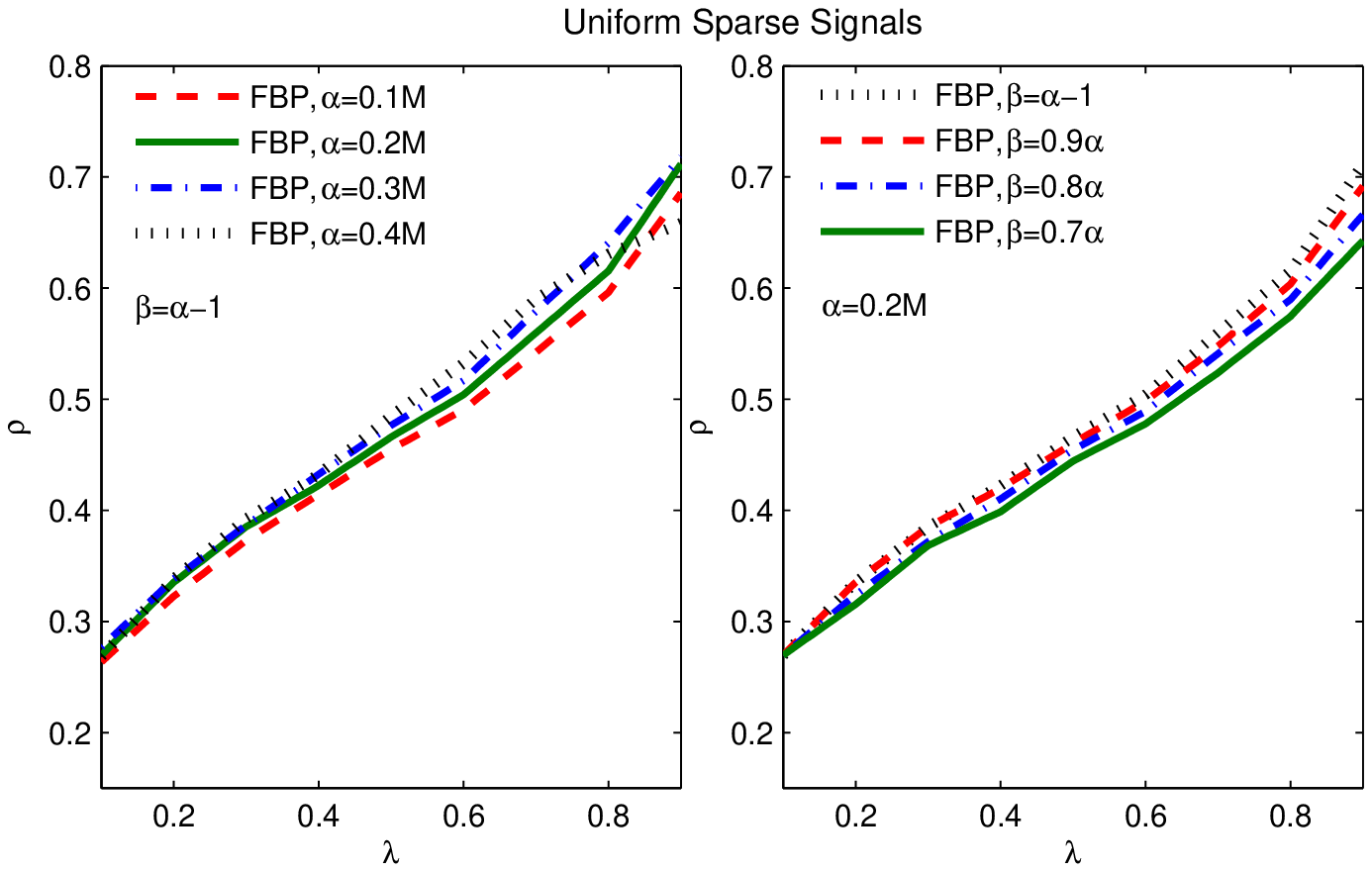}
\includegraphics[width=11.5cm]{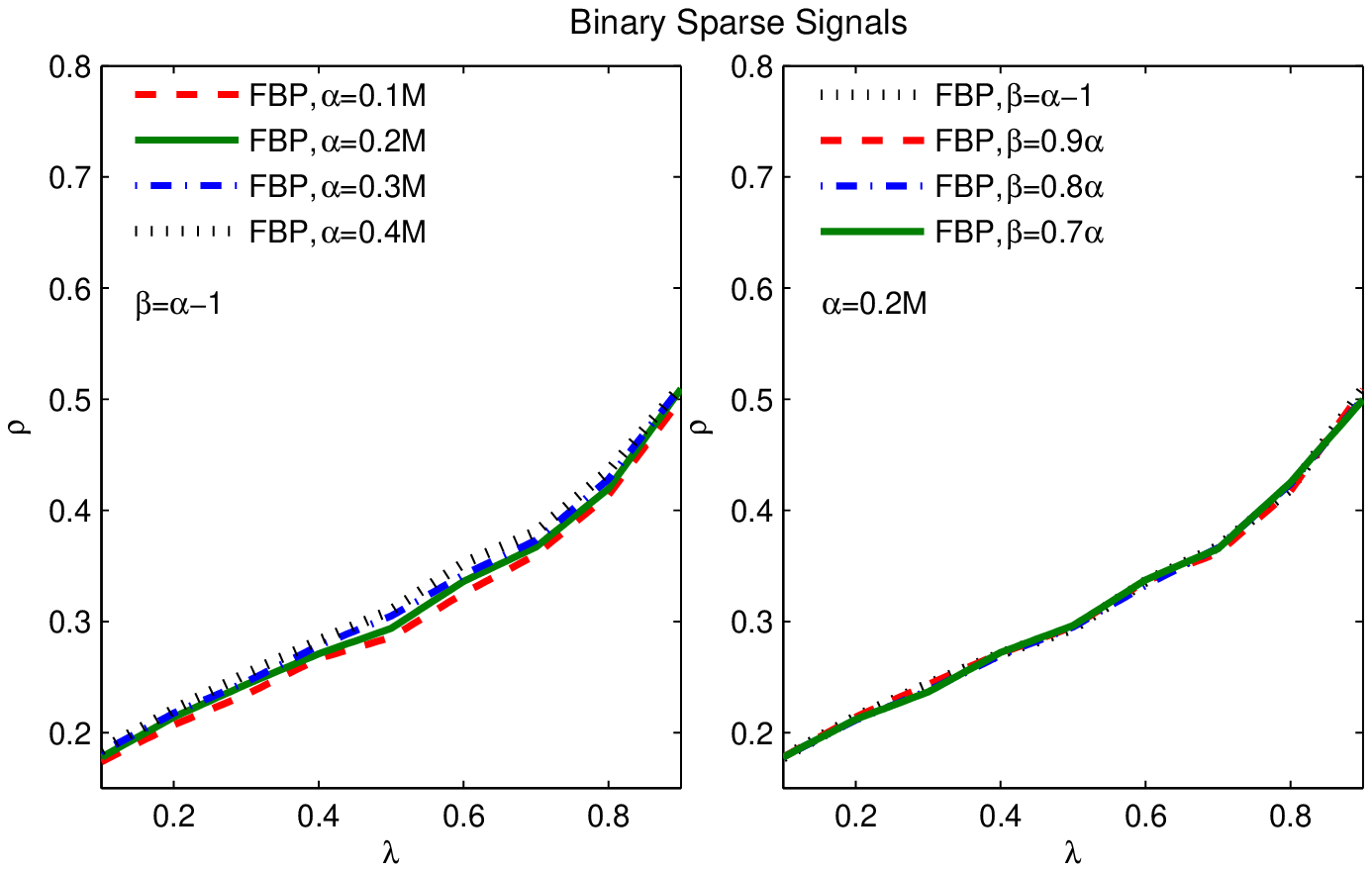}
\caption{Phase transition of FBP with different forward and backward step sizes for the Gaussian, uniform and CARS sparse signals.}
\label{fig:PT__}
\end{center}
\end{figure*}

The phase transitions obtained by the procedure described above are depicted in Figure~\ref{fig:PT__} for the Gaussian, uniform, and binary sparse signals.
These graphs indicate that the performance of FBP fundamentally improves with $\alpha$ and $\beta$, except for very high $\alpha$ choices.
Another exception is the recovery of CARS sparse signals, which constitute the hardest problem for this type of algorithms \cite{Maleki:TST, Dai:SP}. For this case, the gain with $\alpha$ is not significant, while the phase transitions remain unaltered when $\beta$ changes. Another important observation that can be deduced from these results is that the performance of FBP is quite robust to the choice of forward and backward step size choices.

Concentrating on the forward step, the graphs on the left side of Figure~\ref{fig:PT__} reveal that the phase transitions are slightly improved with $\alpha$ until $\alpha=0.3M$ for the uniform and Gaussian sparse signals.
Choosing $\alpha=0.4M$, in contrast, improves the phase transitions only for the mid-$\lambda$ region, while the results get worse especially for the high $\lambda$ values\footnote{We do not increase $\alpha$ over $0.4M$, however note that doing so would even further narrow the mid-$\lambda$ range where the recovery is slightly improved, and widen the high $\lambda$ region where the performance is degraded.}.
The reason for this degradation is that the size of the expanded support estimate exceeds $M$ after the forward step for large $K$ and $\alpha$ values, which leads to an ill-posed orthogonal projection problem, and causes the recovery to fail.
According to Figure~\ref{fig:PT__}, $\alpha=0.3M$ is reasonable for a globally optimum FBP recovery accuracy, while this value might be increased if the problem lies in the mid-$\lambda$ region.
On the other hand, taking into account the computational complexity, we observe no significant decrement in the recovery performance when $\alpha=0.2M$.
Hence, we select $\alpha=0.2M$ below for faster termination, and show that even this choice already leads to better phase transitions than OMP, BP, and SP for the Gaussian and uniform sparse signals\footnote{In fact, the $\alpha$ values evaluated in the previous section do also cover a wide range including the detailed investigation of the choice of $\beta$ for $\alpha=0.2M=20$.}.

As for the backward step, the recovery accuracy decreases slightly with $\beta$ for the Gaussian and uniform sparse signals.
Though this degradation increases slightly with $\lambda$, we observe that the recovery performance of FBP is quite robust to the choice of the backward step size in addition to the forward step size.
Remember that the $\beta/\alpha$ ratio commands the increment in the support size per FBP iteration, and reducing this ratio accelerates the recovery process. Therefore, these results reveal that it is possible to reduce the complexity of FBP by decreasing $\beta/\alpha$. The phase transition comparison below states that the phase transition curves of FBP are still better than those of the BP, SP, and OMP algorithms for the recovery of uniform and Gaussian sparse signals with reduced $\beta/\alpha$ rates.
Similarly, recovery results from the previous section also reveal that FBP does not only provide better recovery rates than the other candidates, but is also as fast as them with $\alpha=20$ and $\beta=17$, which corresponds to $\alpha=0.2M$ and $\beta=0.85\alpha$.

\begin{figure*}[!t]
\begin{center}
\includegraphics[width=\linewidth]{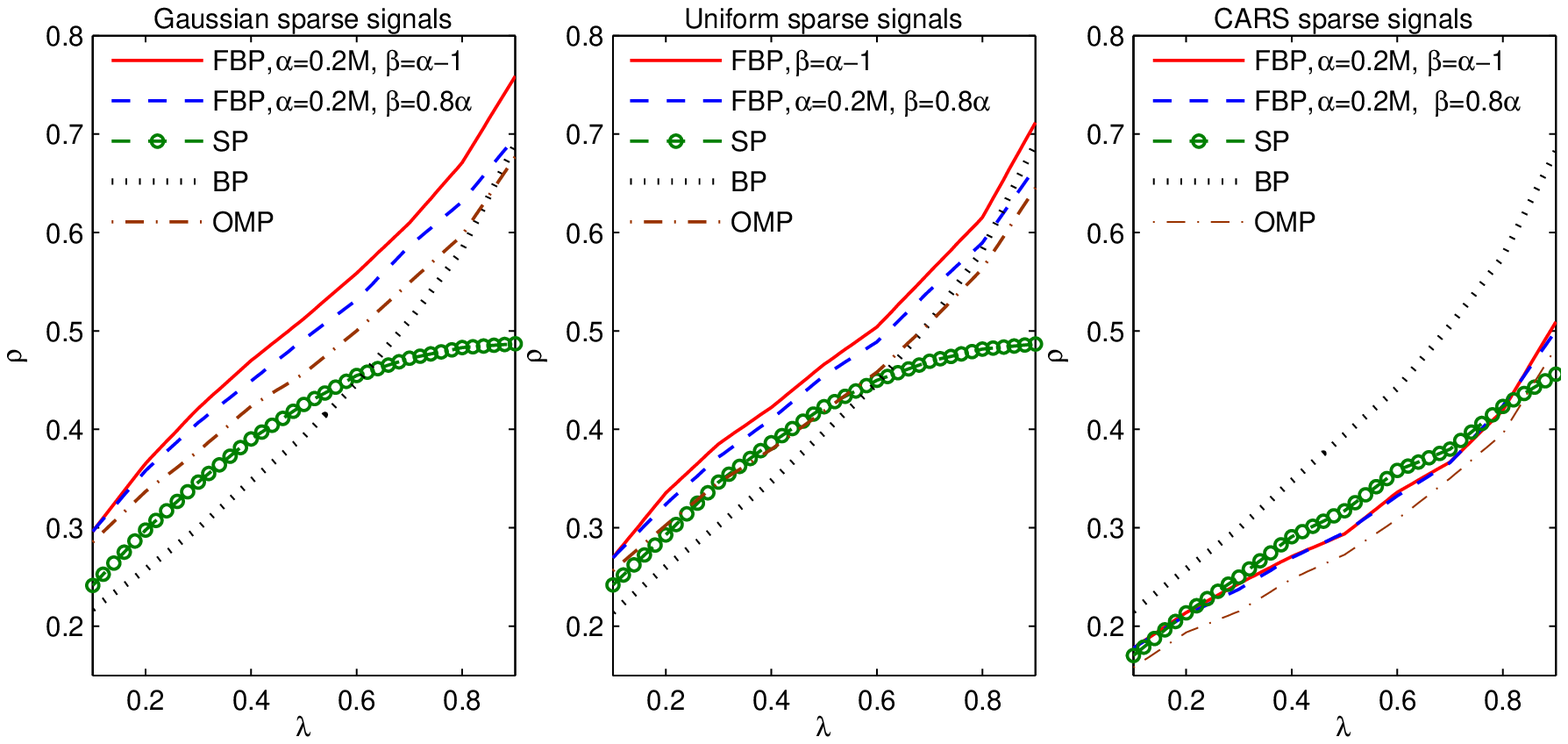}
\caption{Phase transitions of FBP, BP, SP and OMP for the Gaussian, uniform and CARS sparse signals.}
\label{fig:PT}
\end{center}
\end{figure*}

Fig.~\ref{fig:PT} compares the phase transition curve of FBP to those of OMP, BP and SP for the Gaussian, uniform and CARS sparse signals, where FBP is run with both $\alpha=0.2M$, $\beta=\alpha-1$ and $\alpha=0.2M$, $\beta=0.8\alpha$.
For the Gaussian and uniform distributions, FBP outperforms the other algorithms, while for the CARS case BP is better than FBP and the other greedy algorithms.
As a consequence of its strong theoretical guarantees and convex structure, the phase transition of BP is robust to the coefficient distribution.
On the other hand, the performances of the greedy candidates SP, FBP and OMP degrade for the CARS case, while the FBP and OMP curves show the highest variation among different distributions.
We observe that when the nonzero values cover a wide range, as for the Gaussian distribution, the performances of FBP and OMP are boosted.
In contrast, nonzero values of equal magnitudes are the most challenging case for these algorithms. This is related to the involved correlation-maximization step, i.e. choosing the largest magnitude elements of $\mathbf{\Phi}^*\mathbf{r}^{k-1}$, which becomes more prone to errors when the nonzero elements of the underlying sparse signals span a narrower range \cite{MyPhdThesis}.

\begin{figure*}[!t]
\begin{center}
\includegraphics[width=13cm]{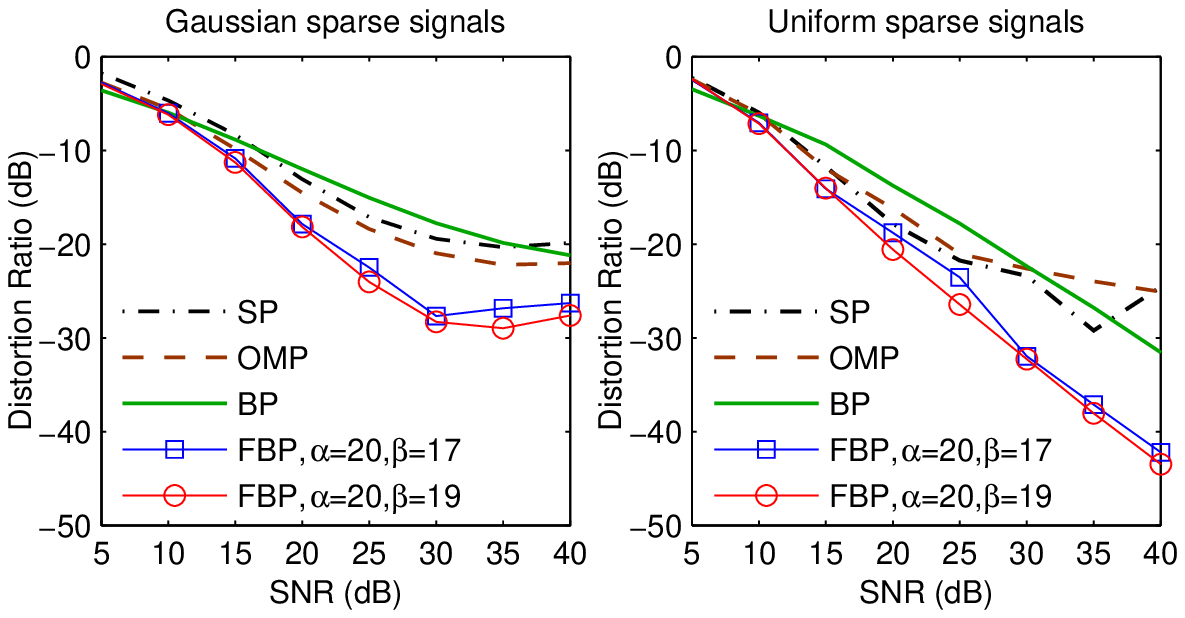}
\end{center}
\caption{Average recovery distortion over SNR in case of noise contaminated observations. For FBP, $\alpha=20$ and $\beta=17$. $K=30$ and $K=25$ for the Gaussian and uniform sparse signals, respectively.}
\label{fig:noisy}
\begin{center}
\includegraphics[width=13cm]{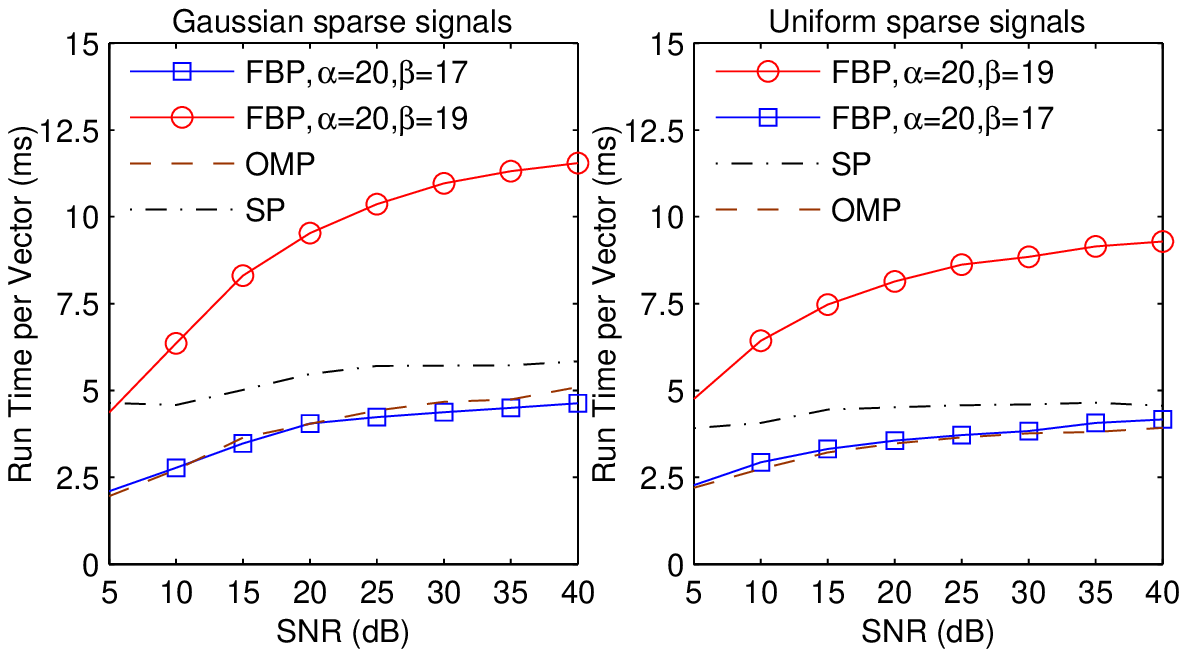}
\end{center}
\caption{Average run time per test sample in case of noise contaminated observations. For FBP, $\alpha=20$ and $\beta=17$. $K=30$ and $K=25$ for the Gaussian and uniform sparse signals, respectively.}
\label{fig:noisytime}
\end{figure*}

\subsection{Recovery from Noisy Observations}

Next, we simulate recovery of sparse signals from noisy observations $\mathbf{y}=\mathbf{\Phi}\mathbf{x} + \mathbf{n}$, which are obtained by contamination of white Gaussian noise component $\mathbf{n}$ at SNR values varying from 5 to 40 dB.
Based on our conclusions from above, FBP is run with both $\alpha=20$, $\beta=19$ and $\alpha=20$, $\beta=17$, which correspond to $\alpha=0.2M$, $\beta=\alpha-1$ and $\alpha=0.2M$, $\beta=0.85\alpha$, respectively.
$\varepsilon$ is selected with respect to the noise level, such that the remaining residual power is equal to the noise power.
The simulation is repeated for 500 Gaussian and 500 uniform sparse signals, where $N=256$ and $M=100$.
The sparsity levels are selected as $K=30$ and $K=25$ for the Gaussian and uniform sparse signals, respectively. $K_{max}$ is 55 as in the first set of simulations. Fig.~\ref{fig:noisy} depicts the recovery error for the noisy Gaussian and uniform sparse signals, while the run times are compared in Fig.~\ref{fig:noisytime}. Note that we express the recovery error in the decibel (dB) scale, calling it the distortion ratio, in order to make it better comparable with SNR. Clearly, FBP yields the most accurate recovery for both $\beta$ values, while BP can do slightly better than FBP only when SNR is 5 dB\footnote{Note that all algorithms almost completely fail at these very low SNR values.}. In addition, we observe that reducing $\beta$ does not significantly change the recovery performance. The run times reveal that FBP is not only the most accurate algorithm in this example, but is also as fast as OMP with $\alpha=20$ and $\beta=17$. As above, this result also supports that $\beta$ can be reduced for speeding up the recovery process without a significant decrement in the recovery accuracy.

\subsection{Demonstration on a Sparse Image}

\begin{figure*}[!t]
\begin{center}
\includegraphics[width=18.2cm]{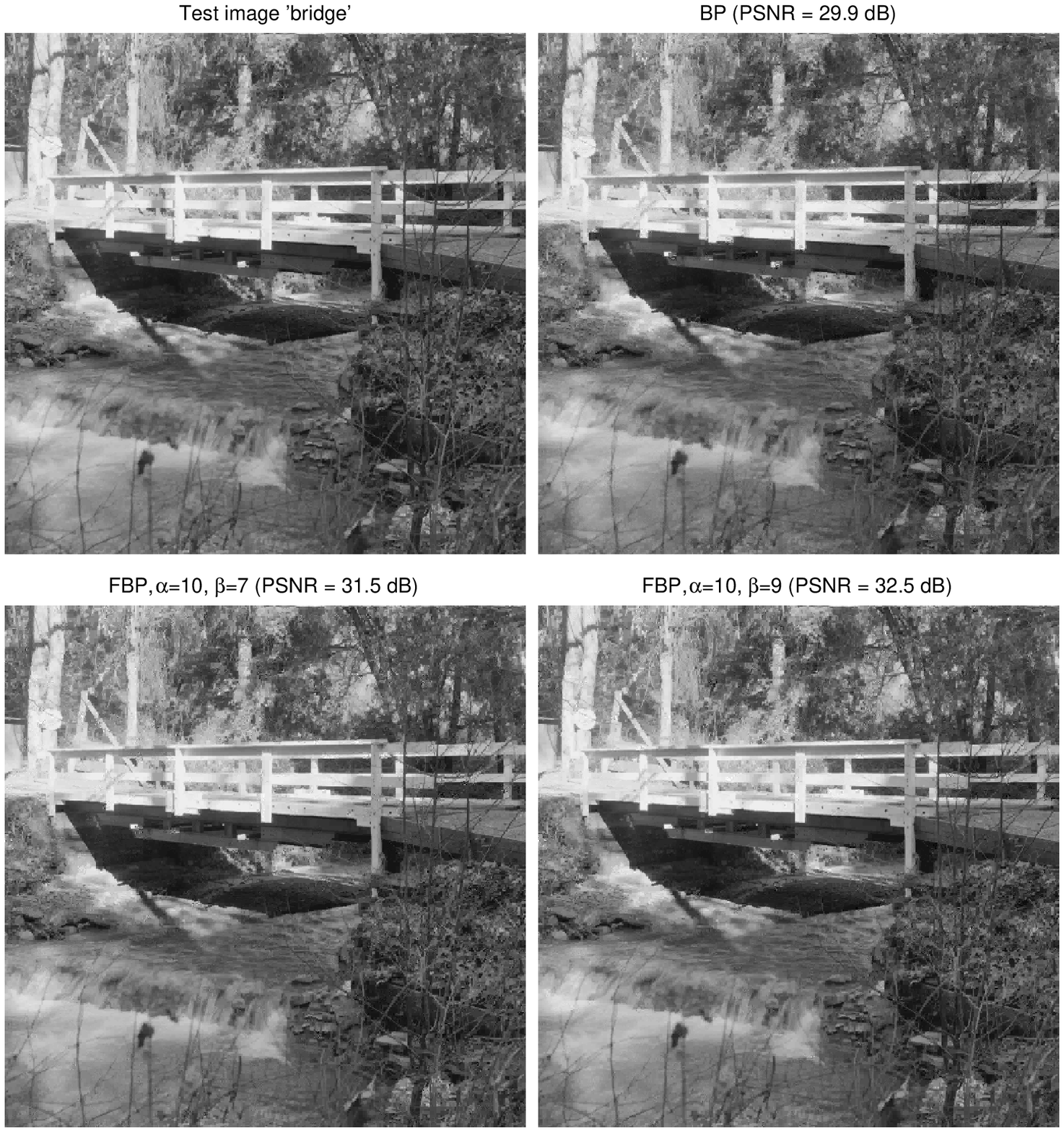}
\end{center}
\caption{Recovery of the image ``bridge'' using BP and FBP. BP recovery yields 29.9 dB PSNR, while FBP provides 31.5 dB PSNR for $\alpha=10$, $\beta=7$ and 32.5 dB PSNR for $\alpha=10$, $\beta=9$.}
\label{fig:bridge}
\end{figure*}

In order to evaluate the FBP recovery performance in a more realistic case, we demonstrate recovery of the $512 \times 512$ image ``bridge''. The recovery is performed using $8 \times 8$ blocks. The aim for such processing is breaking the recovery problem into a number of smaller, and hence simpler, problems. The image ``bridge'' is first preprocessed such that each $8 \times 8$ block is K-sparse in the 2D Haar Wavelet basis, $\mathbf{\Psi}$, where $K=12$, i.e. for each block only the $K=12$ largest magnitude wavelet coefficients are kept. Note that, in this case the signal is not itself sparse, but has a sparse representation in a basis $\mathbf{\Psi}$. Hence, the reconstruction dictionary becomes the holographic basis $\mathbf{V}=\mathbf{\Phi}\mathbf{\Psi}$. From each block, $M=32$ observations are taken, where the entries of $\mathbf{\Phi}$ are randomly drawn from the Gaussian distribution with mean 0 and standard deviation $1/N$. The parameters are selected as $K_{max} = 20$ and $\varepsilon = 10^{-6}$. Two set of FBP parameters are tested, $\alpha=10$, $\beta=7$ and $\alpha=10$, $\beta=9$.
These selections correspond to $\alpha=0.3M$, $\beta=\alpha-1$ and $\alpha=0.3M$, $\beta=0.7\alpha.$\footnote{Since $M$ is small in this case, there is no significant run time difference between choosing $\alpha=0.2M$ and $\alpha=0.3M$. Therefore, we demonstrate FBP with $\alpha=0.3M$. Note that the recovery PSNR that can be obtained with $\alpha=0.2M$ is about 31.5 dB, which is also better than the PSNR value BP yields.}

Fig.~\ref{fig:bridge} shows the preprocessed image ``bridge'' on the upper left. On the upper right is the BP recovery. FBP recovery with $\alpha=10$, $\beta=7$ can be found on lower left, and FBP recovery with $\alpha=10$, $\beta=9$ is next to it. In this example, BP provides a Peak Signal-to-Noise Ratio (PSNR) value of 29.9 dB, while the much simpler FBP improves the recovery PSNR up to 32.5 dB. A careful investigation of the recovered images shows that FBP is able to improve the recovery at detailed regions and boundaries. This example demonstrates that the simpler FBP algorithm is able to perform more accurate and faster recovery of a signal with realistic nonzero coefficient distribution than the much more sophisticated $\ell_1$ norm minimization approach.

\section{Summary}
\label{sec:conc}

This manuscript proposes the forward-backward pursuit algorithm for CS recovery of sparse signals. Falling into the category of TST algorithms, FBP employs a forward step which enlarges the support estimate by $\alpha$ atoms, while the backward step removes $\beta < \alpha$ atoms from it. Hence, this two stage scheme iteratively expands the support estimate for the sparse signal, without requiring $K$ \textit{a priori}, as SP or CoSaMP do.

The presented recovery simulations demonstrate that FBP can provide better exact recovery rates and phase transitions than OMP, SP, and BP except for sparse signals with constant amplitude nonzeros, i.e. CARS ensemble. FBP performance gets better when the magnitudes of the nonzero elements start spanning a wider range, as for the Gaussian distribution. Moreover, FBP is shown to provide a more accurate recovery of a sparse image than the BP algorithm.
Noisy recovery examples state that FBP provides less recovery distortion than OMP, BP and SP for the Gaussian and uniform sparse signals when SNR is greater than 10 dB.
In addition, investigation of the recovery performance with different $\alpha$ and $\beta$ values not only indicates $\alpha\in[0.2M,0.3M]$ and $\beta=\alpha-1$ as a reasonable choice, but also states that shorter forward or backward steps may be incorporated in order to speed up the algorithm with a slight sacrifice of the recovery performance.

Finally, in order to avoid any misinterpretation, we would like to note that our findings do not contradict with the results of Maleki and Donoho in \cite{Maleki:TST}, which mainly investigates the CARS ensemble.
Considering the CARS case only, our findings are parallel to \cite{Maleki:TST}, i.e. SP is the best performer in this worst-case scenario for the greedy algorithms.
However, our results also indicate that FBP provides better recovery than SP and BP when the magnitudes of nonzero elements are not comparable.
This indicates that SP is not the optimum TST scheme for all nonzero element distributions.
Moreover, we believe that only a small portion of the real world problems can be represented with constant amplitude sparse signals. In addition, for most of the problems, it is already known if the signals of interest have comparable magnitudes or not. Accordingly, the performance of algorithms for other distributions should be given more credit. Consequently, we conclude that FBP is a promising algorithm for signal recovery from compressed measurements.

\bibliographystyle{model1-num-names}
\bibliography{FBP_DSP}

\end{document}